# Predictive reinforcement learning based adaptive PID controller

Chaoqun Ma

School of Electronics and Communication Engineerinig, Sun Yat-sen University, Shenzhen, Guangdong, China

Zhiyong Zhang

School of Electronics and Communication Engineerinig, Sun Yat-sen University, Shenzhen, Guangdong, China

## Abstract

**Purpose** – This study aims to address the challenges of controlling unstable and nonlinear systems by proposing an adaptive PID controller based on predictive reinforcement learning (PRL-PID), where the PRL-PID combines the advantages of both data-driven and model-driven approaches.

**Design/methodology/approach** – A predictive reinforcement learning framework is introduced, incorporating action smooth strategy to suppress overshoot and oscillations, and a hierarchical reward function to support training.

**Findings** – Experimental results show that the PRL-PID controller achieves superior stability and tracking accuracy in nonlinear, unstable, and strongly coupled systems, consistently outperforming existing RL-tuned PID methods while maintaining excellent robustness and adaptability across diverse operating conditions.

**Originality/Value** – By adopting predictive learning, the proposed PRL-PID integrates system model priors into data-driven control, enhancing both the control framework's training efficiency and the controller's stability. As a result, PRL-PID provides a balanced blend of model-based and data-driven approaches, delivering robust, high-performance control.

**Keywords:** Predictive reinforcement learning, PID controller, Adaptive control, Model and data driven

## 1. Introduction

In recent years, Artificial Intelligence (AI) has become an essential driver of transformation across various sectors, significantly impacting numerous industries [1]. As a major innovation within the AI technology framework, data-driven control method leverages both offline historical data and real-time operational insights from controlled systems. By combining advanced data processing techniques, this approach facilitates the extraction of system knowledge, enabling precise modeling and control of system dynamics [2,3].

Compared to traditional model-based control methods, data-driven control offers distinct advantages. Firstly, it bypasses the need for extensive system identification or modeling and directly utilizes historical and real-time data to extract relevant information [4]. This minimizes control performance degradation caused by inaccurate or incomplete models, allowing for greater flexibility in addressing system nonlinearity, inaccuracies, and external disturbances. Additionally, data-driven control is more effective in adapting to dynamic environmental changes. By continuously learning and updating from new data, this approach demonstrates superior adaptability and flexibility in complex dynamic environments [5].

While traditional model-based control theories and methods have been extensively researched, many real-world engineering systems are not ideal deterministic systems. The process variables exhibit strong interactivity, and challenges in parameter estimation and disturbance suppression make controlling complex processes extremely difficult [6]. In this context, data-driven control methods emerge as a more flexible and efficient solution.

Neural Networks (NN) and Reinforcement Learning (RL) are the core tools in data-driven control. Neural Networks (NN) are mainly used as approximators, classifiers, or controllers for complex nonlinear systems [7]. The core of RL lies in accumulating experience through interactions with the environment, deriving algorithmic models that lead to optimal decisions [8-10]. Additionally, RL shares a highly similar theoretical framework with traditional closed-loop control. In control problems, the goal is to adjust system inputs such that the system output meets the desired performance criteria [11], this "feedback-adjustment" mechanism is essentially consistent with the structure of RL, where the system (or agent) adjusts the policy (input) to continuously approach the optimal state, achieving better performance in specific tasks. Therefore, RL is frequently applied in control domains for optimizing control strategies [12].

PID controller is the most widely used and fundamental control algorithm in complex process systems. Despite the emergence of advanced control techniques like fuzzy control and adaptive control, the PID controller still dominates industrial control due to its simple structure, ease of implementation, and its ability to effectively ensure control performance under robustness and a wide range of operating conditions [13]. In fact, over 95% of control loops in process control use PID controllers [14], and no other control technology has yet matched the widespread applicability and practical effectiveness of PID controllers [15][14].

Many studies have attempted to integrate the advantages of RL into the design of adaptive PID controllers for various process control applications [6,16-21].[16,17] have proposed adaptive RL-PID controllers for complex nonlinear systems without requiring accurate system models. In [18], the propesed approach combines Proportional-Integral (PI) controllers with RL to optimize electronic throttle valve control, aiming to leverage the stability of the PI controller and the optimization capabilities of RL for efficient control in nonlinear systems. Similarly, another research [21] has developed an automatic PI tuning method based on RL, initially creating a system model using a step-response model, then allowing RL agents to learn from the model in the offline phase to minimize online training time. A recent study [19] proposed a framework for using PID controllers as DRL strategy models, validated in a two-tank level control system. Another method in [6] used RL-based adaptive PID controllers to control both linear and nonlinear unstable systems, employing the improved Proximal Policy Optimization (m-PPO) algorithm to adjust PID parameters online without relying on system models, addressing dynamic uncertainties and nonlinear challenges.

However, traditional RL is inherently a trial-and-error-based control method, where agents interact with the environment and build policy functions through constant trials [22], which often results in inefficient learning. In contrast, infants learn not by engaging in dangerous behavior, but by observing, predicting, and interacting with their environment, which is just the core idea of Model Predictive Control (MPC), a model-based advanced control technique. MPC's primary advantage lies in using a system's dynamic model to make multi-step forward predictions, enabling the system to not only react to current deviations but also proactively avoid potential disturbances and constraint conflicts [23-25]. This model-based prediction capability is the distinguishing feature that sets it apart from trial-and-error learning.

By integrating reinforcement learning (RL) with model-based prediction mechanisms, this paper proposes Predictive Reinforcement Learning-based Adaptive PID Controller (PRL-PID). This controller introduces a reward forecast strategy that enables agents to learn by predicting the system's future dynamics, optimizing control decisions based on predictive information. This strategy combines the strengths of both data-driven and model-driven approaches, allowing agents to focus on both current and future benefits. Additionally, to address issues such as control action abruptness and system delay inertia, action smooth strategy is introduced to enhance dynamic responses, and an improved hierarchical reward function is designed to accelerate training convergence and improve control robustness.

This study aims to propose a dual-driven adaptive controller that retains the flexibility and robustness advantages of data-driven control, while further utilizing known system model priors to help the controller converge more stably and achieve better control performance. To this end, Predictive Reinforcement Learning-based Adaptive PID Controller (PRL-PID) is proposed. In this study, a stable and efficient RL algorithm, Proximal Policy Optimization (PPO) [27], is used to adjust the parameters of the adaptive PID controller. The PRL-PID introduces the reward forecasting strategy, action smooth strategy, and an improved reward function, enabling the controller to achieve fast parameter adaptation in complex dynamic environments. The main contributions of this study include:

- Proposing a predictive RL training method that allows the agent to optimize strategies not only through trial and error but also by utilizing model priors to accelerate the learning process and improve efficiency.
- Introducing a dual-driven adaptive PID controller that optimally utilizes both system input-output data and system dynamics knowledge to optimize control strategies, enabling the controller to maintain flexibility, accuracy, and stability when facing complex dynamic systems.
- Enhancing the controller's efficient convergence and performance improvement through action smooth strategy and the improved reward function.

The remainder of this paper is organized as follows: Section 2 discusses the basic theory of reinforcement learning and the implementation framework of RL-based adaptive PID controller. Section 3 details the specific implementation process of the PRL-PID controller based on predictive reinforcement learning. Section 4 presents simulations and UAV flight control experiments to validate the control performance of the proposed PRL-PID controller in practical applications. Section 5 provides concluding remarks.

## 2. Preliminaries

### *2.1. Reinforcement learning*

Reinforcement Learning (RL) is an algorithmic paradigm where autonomous agents learn optimal behavioral strategies through trial-and-error interactions with dynamic environments. This learning framework operates on a fundamental principle: agents progressively refine their decision-making policies by maximizing cumulative future rewards encoded in environmental feedback signals. The theoretical foundation of RL is established through Markov Decision Processes (MDPs), formally represented as the quintuple $<\mathcal{S}, \mathcal{A}, \mathcal{P}, \mathcal{R}, \gamma>$, where $\mathcal{S}$ denotes the state space, $\mathcal{A}$ represents the action space, $\mathcal{P}$ defines state transition probabilities, $\mathcal{R}$ specifies the reward function, $\gamma \in [0,1)$ is the discount factor for future rewards.

The Markov property requires that the future state depends only on the current state and action, independent of historical trajectories, It also implies that all necessary information for decision-making is contained in the current state $s_t$.

$$P(s_{t+1}|s_t, a_t) = P(s_{t+1}|s_t, a_t, s_{t-1}, \ldots, s_0) \qquad (1)$$

In each step, the agent selects an action $a_t$ based on its current state $s_t$, the environment returns a reward and transitions to the next state $s_{t+1}$. Over continuous repetitions of this loop, the agent refines its policy function to maximize accumulated rewards.

PPO is a widely used actor–critic RL algorithm designed to mitigate instability and low sample efficiency in policy optimization. Compared to traditional policy gradient methods, PPO offers greater stability and sample efficiency, excelling in robotics control, game simulations, and continuous-action tasks [27]. It accomplishes this through the following mechanisms:

- Importance sampling: Reuses experiences generated by past policies, boosting sample efficiency.
- Clipped objective function: Limits the policy-update magnitude to prevent catastrophic policy collapse.

### *2.2. RL-based adaptive PID controller*

RL-based adaptive PID controller dynamically updates control parameters through closed-loop interactions between the agent and the control system. As illustrated in Figure 1, the agent observes the reference signal, the current system output, and any known system dynamics to form its state space. The agent then outputs an action that adjusts the PID parameters $K_p, \tau_i, \tau_d$ in real time. The tuned PID controller calculates system input $u(k)$:

$$u(k) = K_p \left( e(k) + \frac{1}{\tau_i} \sum_{\tau=0}^{k} e(\tau) dt + \tau_d \frac{e(k) - e(k-1)}{dt} \right) \qquad (2)$$

where $K_p, \tau_i$, and $\tau_d$ are updated by the agent's actions, and $e(k)$ is the control error. After saturation limits are applied, the control signal is sent to the plant. By maximizing the reward function, the RL algorithm updates the policy network, forming a closed-loop optimization cycle: (1) state observation, (2) parameter tuning, (3) control execution, (4) reward feedback. Ultimately, this approach achieves adaptive PID tuning and stable convergence in dynamic and complex environments.

During training, the agent attempts to discern the relationship between system inputs and outputs, constructing a mapping from the state space to the optimal control actions. In principle, this allows the PID parameters to be adapted on the fly, dispensing with the conventional reliance on fixed parameters. As a result, such an RL-based adaptive PID controller maintains high control precision and stability even in complex, rapidly changing environments.

Figure 1 Block diagram of RL-adaptive PID controller.

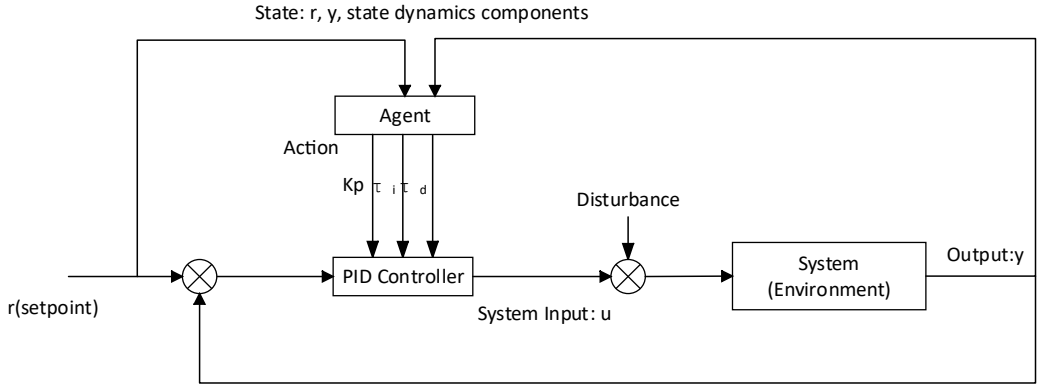

## 3. Predictive RL-Based Adaptive PID Controller

This section details the overall design of the adaptive PID controller based on predictive reinforcement learning. The core algorithm within the PRL-PID controller is a modified PPO framework, integrating reward forecast strategy, action smooth strategy, and hierarchical reward function.

PPO contains a policy network $\pi_\theta(a_t|s_t)$ and a value network $V_\phi(s_t)$. The two networks generate control actions and value-function estimates, respectively, based on the same set of input features. To balance computational efficiency, they share part of the feature extraction layers. The network architecture used in this study is shown in Figure 2, and the parameter settings are listed in Table 1.

For a discretized system, its dynamic model can be described by:
$$x(k+1) = f(x(k), u(k)) \qquad (3)$$

PRL-PID acts as an indirect controller, coupling the policy network with a PID controller. The policy network outputs the tuning parameters of the PID controller, which in turn applies a control signal $u(k)$ to the system. The system then returns a reward $r(k)$ to the agent, which updates its decisions accordingly.

The goal of reinforcement learning is to maximize the expected discounted reward:

$$J(\theta) = E_{\tau \sim \pi_\theta} \left[ \sum_{k=0}^{T} \gamma^k r_k \right] \quad (4)$$

where $r_k$ is the reward at time $k$, $\gamma$ is the discount factor, $\pi_\theta$ denotes the current policy parameterized by θ, and τ is a trajectory generated under $\pi_\theta$. The gradient of the policy network parameters is computed as follows:

$$\nabla_\theta J(\theta) = E_\tau \left[ \sum_{k=0}^{T} \nabla_\theta \log \pi_\theta(a_k|s_k) \cdot \hat{A}_k \right] \quad (5)$$

where $\hat{A}_k$ is the generalized advantage estimator (GAE), measuring how much better the current action is compared to the baseline value function:

$$\hat{A}_k = \sum_{l=0}^{T-k} (\gamma\lambda)^l \delta_{k+l} \quad (6)$$

here $\delta_k$ represents the temporal-difference (TD) error, and $\lambda$ is a smoothing factor used to adjust the GAE calculation process:

$$\delta_k = r_k + \gamma V_\phi(s_{k+1}) - V_\phi(s_k) \quad (7)$$

To enhance training stability, PPO adopts a clipped objective function:

$$L^{CLIP}(\theta) = E_k \left( \min\left( \rho_k(\theta) \hat{A}_k, clip(\rho_k(\theta), 1-\epsilon, 1+\epsilon) \hat{A}_k \right) \right) \quad (8)$$

where $\rho_k(\theta) = \frac{\pi_\theta(a_k|s_k)}{\pi_{\theta old}(a_k|s_k)}$ is the probability ratio between the new policy and the old policy, and $\epsilon$ is a small constant. By restricting changes in the probability ratio, the policy update remains more stable. Meanwhile, the value network is updated to minimize the discrepancy between the predicted and target state values:

$$L^{VF}(\phi) = E_k \left( \frac{1}{2} \left( V_\phi(s_k) - V_k^{target} \right)^2 \right) \quad (9)$$

Figure 2 Actor-Critic network structure.

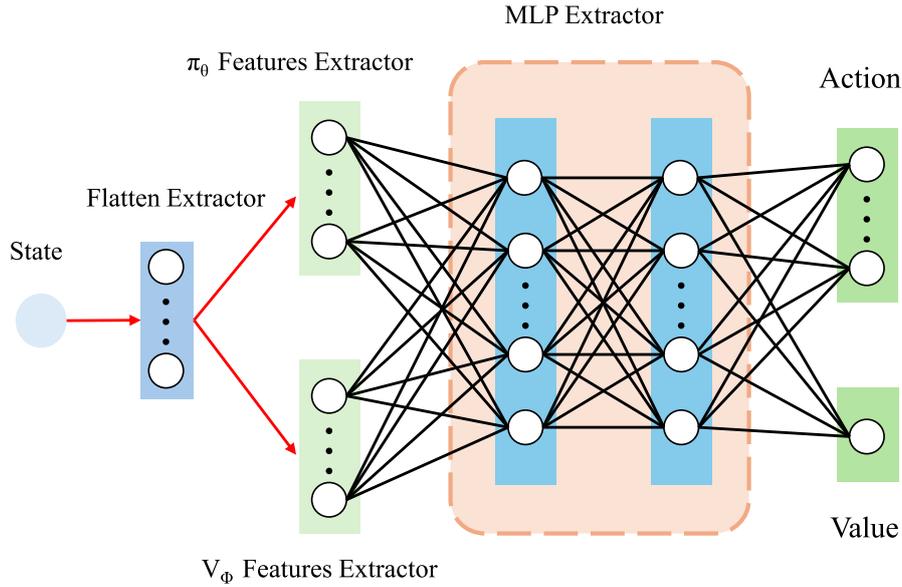

Table 1 Parameter settings for Actor-Critic network.

| Layer Name | Type | Input Features | Output Features | Notes |
| --- | --- | --- | --- | --- |
| State | Input | 4 | | Raw input features. |
| Flatten Extractor | Flatten | | | Reshapes input to a 1D vector. |
| $\pi\ \theta$ Features Extractor | Flatten | | | Feature extraction for policy. |
| $V\Phi$ Features Extractor | Flatten | | | Feature extraction for value network. |
| MLP Extractor 1 | Linear + Tanh | State dimention | 64 | First linear layer. |
| MLP Extractor 2 | Linear + Tanh | 64 | 64 | Second linear layer. |
| Action Network | Linear | 64 | 3 | Output layer for action. |
| Value Output | Linear | 64 | 1 | Output layer for value. |

### *3.1. Hierarchical reward function*

To guide the agent toward efficient learning and precise control in various training phases, a hierarchical reward function is proposed:

$$r_k = -ReLU(\Delta V_k) - \alpha \cdot clip(V_k) - \beta \cdot clip(|e_k|) + r_{add} \qquad (10)$$

where $e_k = r_k - y_k$ is the tracking error, $V_k = \frac{1}{2}e_k^2$ may be the Lyapunov function of the system, $\nabla V_k = V_{k+1} - V_k$ is the differential terms of $V_k$. $r_{add}$ represents the steady-state accuracy incentive.

$$r_{add} = \begin{cases} \frac{0.1}{||e|| + 0.05}, & if ||e|| < 0.05 \\ 0, & otherwise \end{cases} \quad (11)$$

The $-ReLU(\Delta V_k)$ term helps penalize rapid changes in the error, preventing the system from diverging during the initial training phase. Additional terms $-\alpha \cdot clip(V_k)$ and $-\beta \cdot clip(|e_k|)$ can be introduced to address different stages of training. Moreover, a supplemental reward term $r_{add}$ can be applied when the absolute error is smaller than a preset threshold, providing fine-grained guidance for reducing residual error and achieving high-accuracy tracking.

### 3.2. Reward forecast strategy

The reward forecast strategy in PRL-PID enables the agent to learn by predicting future system states. Specifically, given the current system state $s_k$ and PID parameters, the future states over $N$ steps are recursively estimated using the system's prior model:

$$\begin{aligned}\hat{x}(k+1) &= f(x(k), u(k)) \\ \hat{x}(k+2) &= f(\hat{x}(k+1), \hat{u}(k+1)) \\ &\vdots \\ \hat{x}(k+N) &= f(\hat{x}(k+N-1), \hat{u}(k+N-1))\end{aligned} \quad (12)$$

where, for simplicity and stability, the control input within these $N$ steps is generated by a fixed set of PID parameters:

$$\hat{u}(k+i) = K_p^{(k)}\hat{e}(k+i) + K_i^{(k)}\sum_{\tau=0}^{k+i}\hat{e}(\tau)dt + K_d^{(k)}\frac{\hat{e}(k+i) - \hat{e}(k+i-1)}{dt} \quad (13)$$

After executing action $a_k$ a in state $s_k$, the next $N - step$ trajectory is predicted, and an average reward is computed:

$$r_k^{pred} = \frac{1}{N}\sum_{i=0}^{N-1} r_{k+i} \quad (14)$$

Here, $r_{k+i}$ is the immediate reward at step $k + i$, calculated by the reward function $R(\hat{s}_{k+i}, \hat{a}_{k+i})$. The predicted reward $r_k^{pred}$ replaces the original single-step reward, allowing the agent to incorporate model priors for more efficient learning. Thus, a data-and-model-driven design is realized, combining the flexibility of data-driven methods with the foresight offered by model-based approaches.

### 3.3. Action smooth strategy

To further optimize the agent's control decisions, action smooth strategy is adopted. In standard RL, the agent typically executes an action immediately at each time step. However, because of system response lag and inertia, these instantaneous actions can be too short-lived

---

**Algorithm 1** PRL-PID Controller Training

---
1: Initialize policy parameters $\theta$, value network parameters $\phi$, discount factor $\gamma = 0.99$, batch size $B = 256$, lookahead steps $N$.
2: **for** $k = 0$ to $K_{max}$ **do**
3:    **Collect trajectories:**
4:    for each trajectory, perform steps 5 to 14
5:    **for** $t = 0$ to $T$ **do**
6:       Select action based on policy $\pi_\theta$ and execute action smooth strategy:

$$\hat{a}_t = [Kp, \tau_i, \tau_d]$$

7:       **for** $i = 0$ to $N - 1$ **do**
8:          Compute control signal using PID:

$$\hat{u}_{t+i} = K_p \left( e_t + \frac{1}{\tau_i} \sum_{\tau=0}^{k+i} \hat{e}(\tau) \, dt + \tau_d \frac{\hat{e}(k+i) - \hat{e}(k+i-1)}{dt} \right)$$

9:          Predict future state $\hat{s}_{t+i}$ and calculate the reward $\hat{r}_{t+i}$
10:       **end for**
11:       Perform lookahead reward calculation:

$$r_t = \frac{1}{N} \sum_{i=0}^{N-1} \hat{r}_{t+i}$$

12:       Execute $u_t$ into system and receive the next state $s_{t+1}$
13:       Store $(s_t, a_t, r_t, s_{t+1})$ in experience buffer
14:    **end for**
15:    **Compute advantage function:**

$$\delta_t = r_t + \gamma V_\phi(s_{t+1}) - V_\phi(s_t), \quad \hat{A}_t = \sum_{l=0}^{\infty} (\gamma \lambda)^l \delta_{t+l}$$

16:    **Update policy network:**
17:    **for** $epoch = 1$ to $10$ **do**
18:       Sample minibatch $\{s_t, a_t, \hat{A}_t\}$ from experience buffer
19:       Compute importance sampling ratio $\rho_t(\theta)$
20:       Update policy parameters:

$$\theta \leftarrow \arg\max_\theta \frac{1}{B} \sum_{t=0}^{T} \min\left( \rho_t \hat{A}_t, \ \text{clip}(\rho_t, 1 \pm \epsilon)\hat{A}_t \right)$$

21:    **end for**
22:    **Update value network:**
23:    Compute target value $V_{target}$
24:    Update value network parameters:

$$\phi \leftarrow \arg\min_\phi \frac{1}{B} \sum_{t=0}^{T} \left( V_\phi(s_t) - V_{target} \right)^2$$

25: **end for**

to meaningfully influence system dynamics. In fact, if consecutive actions differ drastically, the effect of earlier actions may be overshadowed or completely nullified, preventing the agent from learning valuable experience from those actions.

To address this issue, the proposed strategy retains a portion of the historical action information to bolster the dynamic response of the system. Concretely, at each time step $t$, the agent's executed action is not the newly generated action alone but instead a weighted combination of the past $M$ actions:

$$\hat{a}(t) = \sum_{i=t-M+1}^{t} w_{t,i} \cdot a(i) \tag{15}$$

where $\hat{a}(t)$ is the action actually executed at time $t$, $a(i)$ is the immediate action generated at time $t - i$, and $w_{t,i}$ are weighting coefficients controlling the contribution of historical actions.

By preserving part of the historical input, the system's state evolution accumulates more effectively over time, ensuring that the current action exerts a longer-lasting influence on system behavior. Moreover, action smooth helps curb abrupt spikes in the control signal, thereby mitigating undesirable oscillations, overshoot, or even nonlinear instabilities—problems especially unacceptable in applications requiring precise or highly reliable control. Consequently, action smooth significantly reduces the risk of instability arising from abrupt control updates.

Finally, the complete algorithm for the predictive RL-based adaptive PID controller is shown in Algorithm 1.

## 4. Performance validation of PRL-PID controller

To verify the effectiveness and robustness of PRL-PID controller in handling unstable system dynamics, we performed simulations on linear unstable systems with various dynamic characteristics, and further demonstrated the applicability of the PRL-PID controller under diverse operating scenarios. The training framework was implemented in Python. During simulation, the known dynamic properties of each system were discretized to accommodate the discrete control and state spaces commonly used in reinforcement learning. The sampling time was set to $T_s = 0.1s$. Throughout training, the reference setpoint was randomly varied within specified upper and lower limits to emulate setpoint changes that might occur under different operational conditions. The objective was to adapt the PID parameters $K_p, \tau_i, \tau_d$ in different system states and environments, thereby achieving stable tracking of the specified setpoint. To ensure a fair comparison, we used the same 500,000 training iterations as the standard for all experiments, allowing each control algorithm to optimize its strategy under identical training durations. This setup provided a consistent basis for subsequent performance comparisons.

### *4.1. Linear Unstable System*

In this study, a representative first-order unstable system was used to validate the PRL-PID controller. The system contains an unstable pole at $p = 1$, and its transfer function is given by:

$$G(s) = \frac{1}{s-1} \quad (16)$$

The control performance of PRL-PID controller was compared with that of three other RL-based PID controllers [27-29]. Figure 3 presents the step-response curves for each controller, with tuning performed by four different RL algorithms. The SAC-PID controller underwent oscillatory action updates that led to a second divergence, and its output was unable to stabilize. In contrast, the other three controllers successfully tracked the specified setpoint. Among these, the TD3-PID controller exhibited the largest overshoot, while the PPO-PID controller took the longest to settle. The PRL-PID controller achieved favorable results on both metrics and, thanks to the action smooth strategy, demonstrated lower system oscillations.

Figure 3 Servo response of $G(s) = \frac{1}{s-1}$.

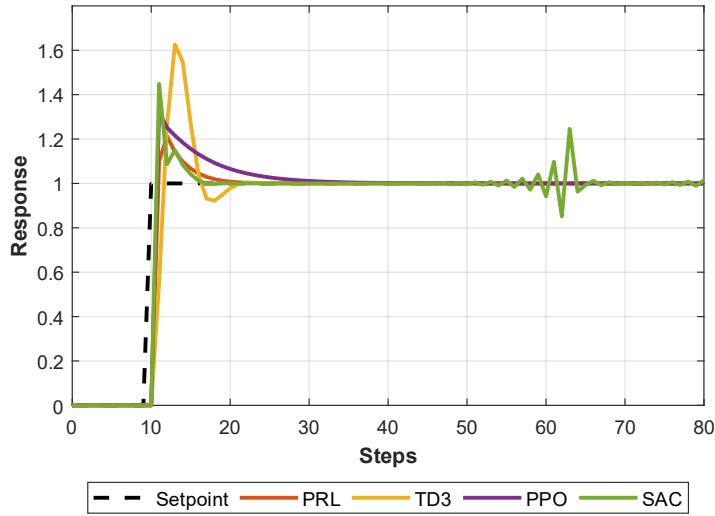

### *4.2. Effectiveness of the reward forecast strategy*

To further test the effectiveness of the reward forecast strategy, this study selected a classic second-order unstable system for simulation [30]. The transfer function is given by:

$$G(s) = \frac{-0.2679(1 - 41.667s)}{279.03s^2 - 2.9781s + 1} \tag{17}$$

This system includes two conjugate poles with positive real parts $0.0053 \pm 0.0596i$ and one zero in the right-half plane $z = 0.024$, underscoring its substantial instability and posing significant challenges for optimizing the agent's control policy. Figure 4 illustrates how the reward forecast strategy influences the training reward. Compared to a purely trial-and-error learning approach, incorporating forward-looking rewards yielded faster reward growth and higher overall reward levels. Among different values of $N$, $N = 3$ offered the best control performance. Figure 5 also shows the step responses under various forward-looking horizon lengths, revealing that when $N = 3$, the controller achieved the smallest overshoot and settling time. These results indicate that introducing the reward forecast strategy not only accelerated controller convergence during training but also improved dynamic performance in real-time operation.

Figure 4

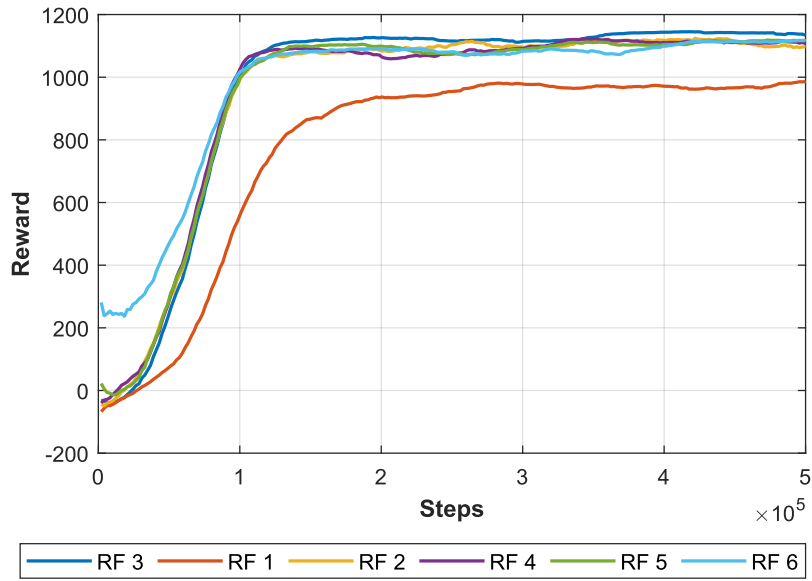

Figure 5 Effect of reward forecast strategy.

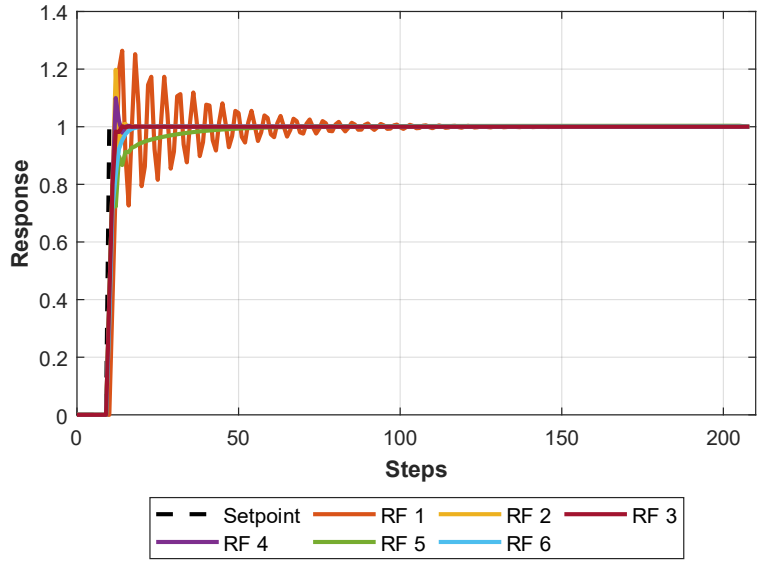

### *4.3. Action smooth strategy*

The action smooth strategy was introduced to address problems stemming from system delay, inertia, and other factors that restrict the impact of instantaneous actions on system dynamics and may lead to output oscillations. Such issues are especially pronounced in certain unstable systems. To investigate how the smoothing mechanism affects controller performance, a representative first-order unstable system was selected for simulation. Its transfer function is:

$$G(s) = \frac{1}{s - 1.5} \tag{18}$$

This system features a single pole at $p = 1.5$ in the right-half plane, causing exponentially divergent behavior that imposes strict timing precision on the control inputs.

Three smoothing strategies are compared in this study:

1. Simple Moving Average (SMA):

$$a_t^{smooth} = \frac{1}{M} \sum_{k=0}^{M-1} a_{t-k}^{raw}, \quad t \geq M \tag{19}$$

where $a_t^{raw}$ is the original action, and $M$ is the sliding window length.

2. Linearly Weighted Moving Average (LWMA):

$$\begin{cases} w_k = \alpha + \dfrac{(1-\alpha)(k+1)}{M} \\ a_t^{smooth} = \dfrac{\sum_{k=0}^{M-1} w_k a_{t-k}^{raw}}{\sum_{K=0}^{M-1} w_k} \end{cases} \quad (20)$$

where $\alpha = 0.2$ is the weight-decay factor.

3. Exponential Recursive Average (ERA):

$$a_t^{smooth} = \begin{cases} a_0^{raw}, & t = 0 \\ \alpha a_t^{raw} + (1-\alpha) a_{t-1}^{smooth}, & t > 0 \end{cases} \quad (21)$$

where $\alpha = 0.3$ is the discount factor.

All three smoothing strategies effectively mitigated large and abrupt shifts in the control signal by retaining historical action information, thereby significantly enhancing system stability. Initially, the raw action signals produced marked fluctuations, especially at the beginning of the system's response. Simple moving average (SMA) and Linear weighted moving average (LWMA 5) successfully filtered high-frequency oscillations but introduced more pronounced delay in the action response. By contrast, exponential recursive average (EMA 5) simultaneously preserved rapid response and short settling time while reducing system overshoot, thus stabilizing the control process more effectively.

Figure 6 Servo response of different smooth strategies.

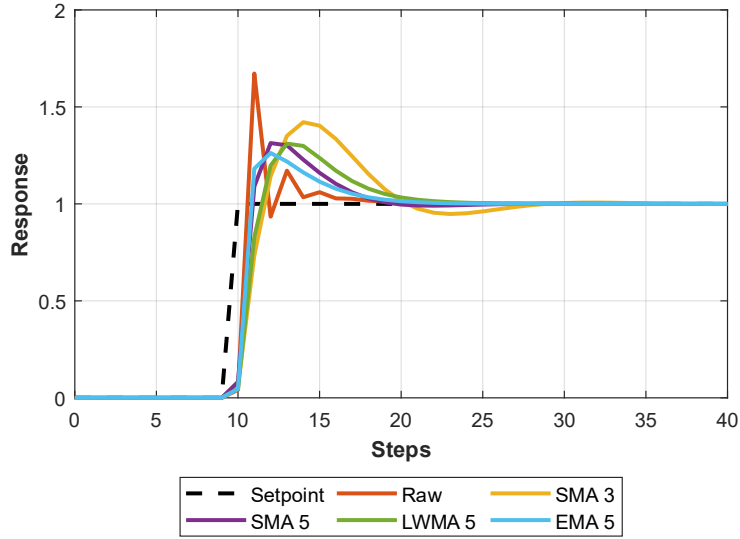

### 4.4. Time-Varying system

To further evaluate the adaptability and robustness of PRL-PID controller in dynamic environments, we apply it to a first-order plant with explicitly time-varying dynamics:

$$G(s,t) = \dfrac{1}{s - a(t)} \quad (22)$$

where the time-varying parameter is defined as:
$$a(t) = 0.5 + 0.25(\sin(0.2\pi t) + 1) \tag{23}$$

During the servo-regulation experiments, the agent is trained using reference values only within the interval [0,2], whereas in the validation phase the controller is required to track targets over an extended range of [−0.3,2.5],. Figure 7 compares the regulation responses obtained under different control strategies. When neither the reward-foresight (RF) nor the action-smoothing (AS) strategy is employed, the PPO-PID controller exhibits a pronounced offset for reference values in [1.5,2.5],. Introducing the AS strategy alone substantially reduces this offset. When only the RF strategy is applied, the controller tracks stably throughout the range except at the extreme reference of 2.5, where a noticeable offset remains. Finally, when both RF and AS strategies are combined, the controller achieves accurate tracking within the training range and maintains strong generalization performance even under operating conditions beyond that range.

Figure 7 Regulation responses of the time-varying first-order system under different control strategies.

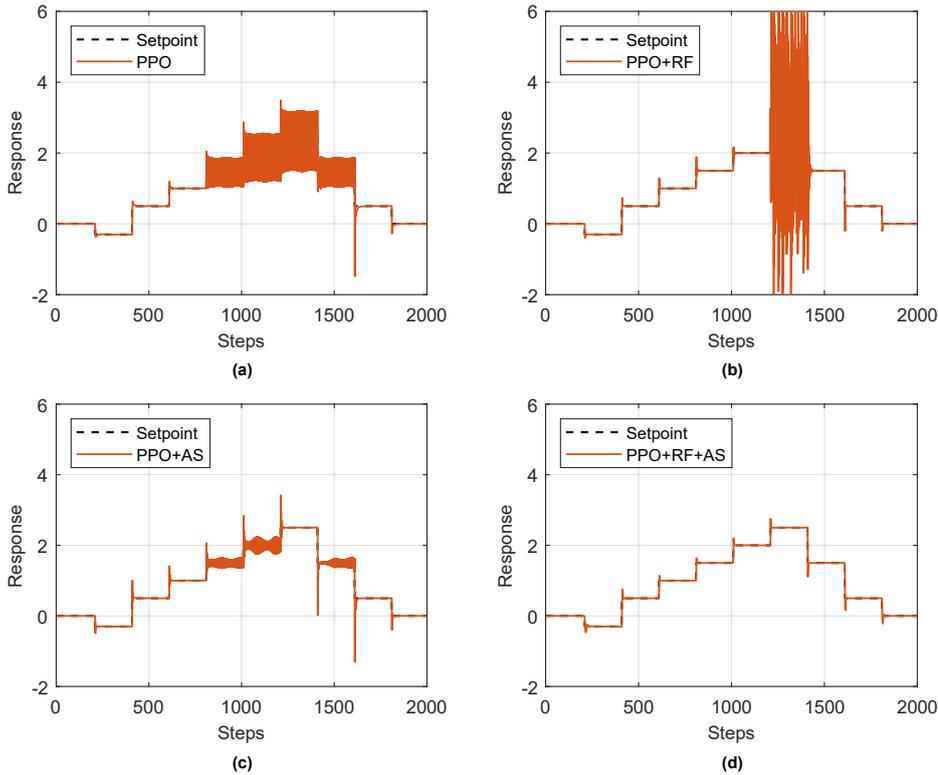

### 4.5. Nonelinear system

The two-tank control system is a prototypical nonlinear, multivariable coupled process in the field of process control. Its basic configuration is shown in Figure 8. The system consists of two tanks connected in series: the inlet flow $q_{in}$ first enters the upper tank (Tank 1). Once the liquid level $h_1$ in Tank 1 reaches a certain height, the hydrostatic pressure drives the fluid through a valve into the lower tank (Tank 2). By varying the inlet flow rate $q_{in}$ in to the upper tank, one can regulate the liquid level $h_2$ in the lower tank.

The dynamic behavior of the two-tank system can be described by the following set of nonlinear differential equations:

$$\begin{cases} \dot{h}_1 = \frac{1}{A_1}(q_{in} - k_1\sqrt{\max(h_1 - h_2, 0)}) \\ \dot{h}_2 = \frac{1}{A_2}(k_1\sqrt{\max(h_1 - h_2, 0)} - k_2\sqrt{\max(h_2, 0)}) \end{cases} \quad (24)$$

where $h_1$ and $h_2$ denote the liquid levels in the upper and lower tanks, respectively, $A_1$ and $A_2$ are their cross-sectional areas, $q_1$ is the externally adjustable inlet flow rate, and $k_1$ and $k_2$ are the valve flow coefficients. To emulate realistic industrial conditions, zero-mean Gaussian noise is superimposed on the flow measurements and actuator commands, and both pulse and step disturbances are introduced after the system has reached steady state—these simulate instantaneous flow shocks and sudden changes in valve characteristics.

Figure 8 Two-tank control system.

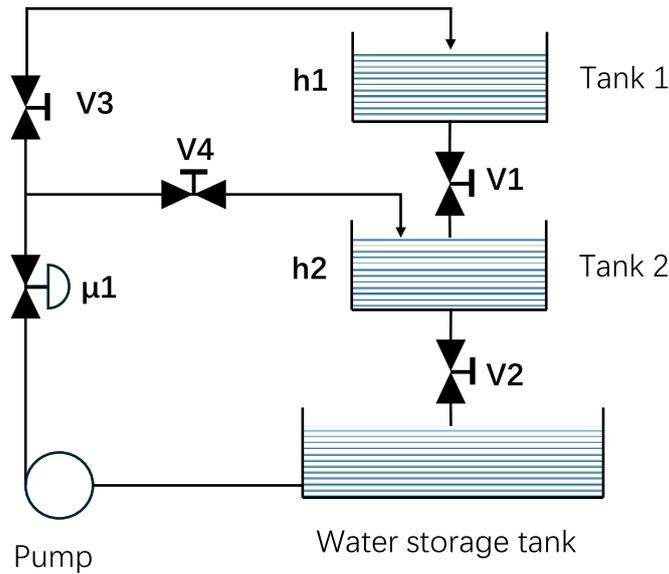

Figure 9 presents the servo-response of the two-tank system under various control strategies: negative and positive pulse disturbances are applied at $k = 100$ and $k = 300$, respectively, and a step disturbance is imposed at $k = 500$. The Integral square error (ISE) and Integral Absolute Error (IAE) are shown in Table 2 .The results demonstrate that the proposed reward-foresight strategy significantly reduces the settling time, while the action-smoothing strategy further diminishes both the steady-state error and the overall regulation time without compromising input stability.

Table 2 Error metrics of different controllers.

| Sampling interval | PPO | | PPO+RF | | PPO+AS | | PPO+RF+AS | |
| --- | --- | --- | --- | --- | --- | --- | --- | --- |
| | ISE | IAE | ISE | IAE | ISE | IAE | ISE | IAE |
| 0≤k≤200 | 1.163 | 4.094 | 0.437 | 1.642 | 1.056 | 3.639 | 0.311 | 1.240 |
| 200≤k≤400 | 8.572 | 12.981 | 3.213 | 5.176 | 8.423 | 12.707 | 2.002 | 3.727 |
| 400≤k≤600 | 5.597 | 10.770 | 2.359 | 4.784 | 5.621 | 10.975 | 1.884 | 3.963 |
| 600≤k≤800 | 0.174 | 1.598 | 0.064 | 0.789 | 0.203 | 1.959 | 0.020 | 0.462 |

Figure 9 Servo Response of the Two-Tank Liquid-Level System under Different Control Strategies.

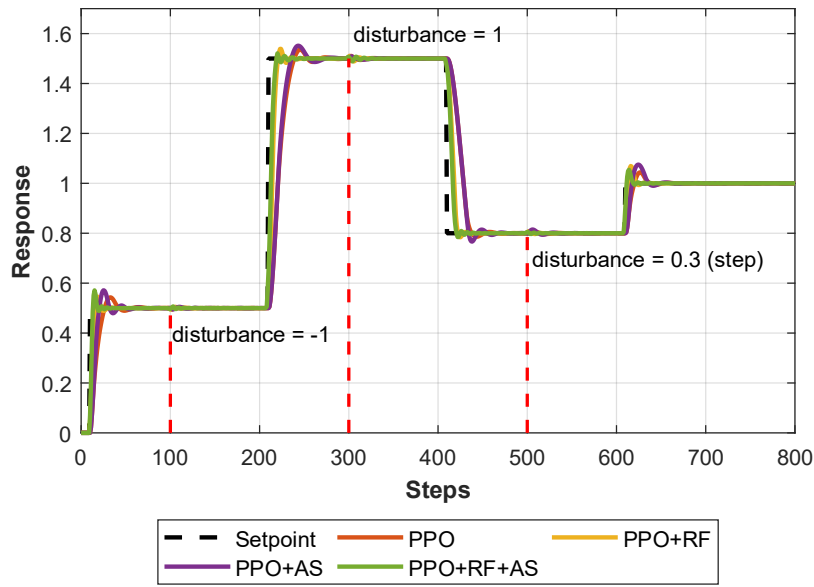

### *4.6. UAV control system*

In recent decades, researchers have increasingly directed their attention toward the various types of unmanned aerial vehicles (UAVs), which exhibits highly coupled rotational dynamics [31]. A MIMO attitude-control simulation environment of UAV comprising three degrees of freedom(DOF) including roll($\phi$), pitch($\theta$) and yaw($\psi$) is constructed. Each DOF is regulated by an independent PID controller. In this study, The classical simplified attitude dynamics model of a quadrotor is employed to capture the nonlinear coupling among these three channels:

$$\begin{cases} \ddot{\phi} = \dfrac{I_y - I_z}{I_x}\dot{\theta}\dot{\psi} + \dfrac{1}{I_x}U_\phi \\ \ddot{\theta} = \dfrac{I_z - I_x}{I_y}\dot{\phi}\dot{\psi} + \dfrac{1}{I_y}U_\theta \\ \ddot{\psi} = \dfrac{I_x - I_y}{I_z}\dot{\phi}\dot{\theta} + \dfrac{1}{I_z}U_\psi \end{cases} \quad (25)$$

where $U_\phi, U_\theta$ and $U_\psi$ denote the control torques about the roll, pitch and yaw axes, respectively, and the principal moments of inertia are $I_x = 0.4, I_y = 0.4$ and $I_z = 0.8$.

To address the deterioration in control performance caused by the dynamic coupling among roll, pitch, and yaw attitude channels during quadrotor UAV flight, this study proposes a cross-axis coupled PID control scheme built upon the classical PID controller. Specifically, in the control input of each channel, error feedback information from the other two channels is introduced. The form of cross-axis PID control inputs can be expressed as follows:

$$\begin{cases} U_\phi(k) = K_{p,\phi}\left(e_\phi^{couple}(k) + \dfrac{1}{\tau_{i,\phi}}\sum_{\tau=0}^{k}e_\phi^{couple}(\tau)dt + \tau_{d,\phi}\dfrac{e_\phi^{couple}(k) - e_\phi^{couple}(k-1)}{dt}\right) \\ U_\theta(k) = K_{p,\theta}\left(e_\theta^{couple}(k) + \dfrac{1}{\tau_{i,\theta}}\sum_{\tau=0}^{k}e_\theta^{couple}(\tau)dt + \tau_{d,\theta}\dfrac{e_\theta^{couple}(k) - e_\theta^{couple}(k-1)}{dt}\right) \\ U_\psi(k) = K_{p,\psi}\left(e_\psi^{couple}(k) + \dfrac{1}{\tau_{i,\psi}}\sum_{\tau=0}^{k}e_\psi^{couple}(\tau)dt + \tau_{d,\psi}\dfrac{e_\psi^{couple}(k) - e_\psi^{couple}(k-1)}{dt}\right) \end{cases} \quad (26)$$

where the cross-axis coupled error inputs are given by:

$$\begin{cases} e_\phi^{couple} = (\phi_{ref} - \phi) + c_{rp}\theta + c_{ry}\psi \\ e_\theta^{couple} = (\theta_{ref} - \theta) + c_{pr}\phi + c_{py}\psi \\ e_\psi^{couple} = (\psi_{ref} - \psi) + c_{yr}\phi + c_{yp}\theta \end{cases} \quad (27)$$

Here, $\phi_{ref}, \theta_{ref}, \psi_{ref}$ denote the target setpoints for roll, pitch, and yaw angles, respectively, $\phi, \theta, \psi$ represent the actual measured angles, and the parameters $c_{rp}, c_{ry}, c_{pr}, c_{yp}, c_{yr}, c_{yp}$ are cross-axis coupling coefficients. The specific values of these coupling coefficients are dynamically optimized online through reinforcement learning algorithm to adaptively accommodate the coupling characteristics of the system.

Figure 10 presents the UAV attitude-control simulation results obtained with the cross-axis-coupled PID scheme. Even under multi-input, multi-output conditions with highly coupled, nonlinear attitude dynamics, the PRL-PID controller tracks the desired attitude angles rapidly and accurately, demonstrating excellent stability and robustness.

Figure 10 Attitude control of UAV using cross-axis coupled PID control scheme.

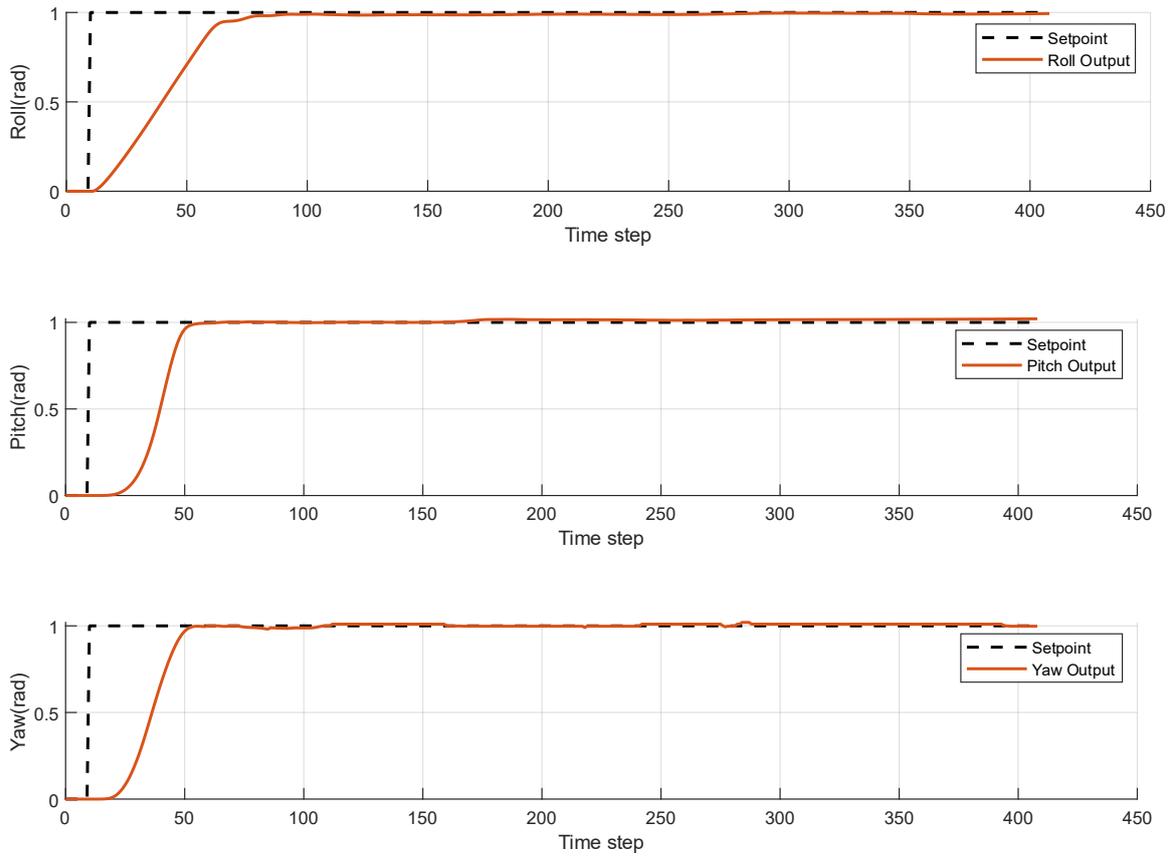

## *5. Conclusion*

The predictive RL-based adaptive PID controller (PRL-PID) proposed in this study offers an efficient and flexible control solution by integrating both data-driven and model-based insights. Building on conventional reinforcement learning training algorithms, PRL-PID incorporates a reward forecast strategy that enables the agent to leverage system-model priors for predictive learning, thereby accelerating convergence and substantially enhancing controller performance. In addition, action smooth strategy and hierarchical reward function have been introduced to further strengthen the controller's stability and convergence. Comparative simulations on linear unstable, time-varying, and highly coupled nonlinear processes, as well as a simplified quadrotor attitude model, consistently showed that PRL-PID reduces overshoot, shortens settling time, and maintains strong robustness against disturbances.